\documentclass[3p]{elsarticle}

\usepackage{lineno,hyperref}
\usepackage{xspace}
\modulolinenumbers[1]
\usepackage{bm}
\usepackage{amssymb}
\usepackage{hyperref}
\usepackage{amsmath}
\usepackage{graphicx}
\usepackage{comment}
\usepackage{color}

\newcommand{\s}{\ensuremath{\sqrt{s}}\xspace}
\newcommand{\ups}{\ensuremath{\Upsilon}(nS)\xspace}

\newcommand{\meanptups}{\ensuremath{\langle p_{\mathrm{T}}^{\Upsilon} \rangle}\xspace}
\newcommand{\meanptjet}{\ensuremath{\langle p_{\mathrm{T}}^{\mathrm{J}} \rangle}\xspace}
\newcommand{\gevc}{\ensuremath{\rm{GeV/}c}\xspace}
\newcommand{\pt}{\ensuremath{p_{\mathrm{T}}}\xspace}
\newcommand{\qqbar}{\ensuremath{Q\bar{Q}}\xspace}
\newcommand{\ntrack}{\ensuremath{N_{\mathrm{Track}}}\xspace}
\newcommand{\py}{Pythia~8.312\xspace}

\journal{Journal of \LaTeX\ Templates}

\begin{document}

\begin{frontmatter}

\title{{\bf Multiplicity dependence of $\Upsilon$(nS) mean transverse momentum in proton-proton collisions}}

\author{Luis Gabriel Gallegos Mariñez, Lizardo Valencia Palomo* and Luis Cedillo Barrera}
\address{Departamento de Investigaci\'on en F\'isica, Universidad de Sonora, \\ Blvd. Luis Encinas y Rosales S/N, Col. Centro, Hermosillo, Sonora, M\'exico}

\cortext[mycorrespondingauthor]{lizardo.valencia@unison.mx}

\begin{abstract}

Correct description of quarkonia production and kinematics are still one of the most challenging assignments for Quantum Chromodynamics. This document presents a study of the $\Upsilon$(1S), (2S) and (3S) mean transverse momentum (\meanptups) as a function of the charged particle multiplicity (\ntrack) in proton-proton collisions at \s{} = 7 TeV generated with \py{} CUETP8M1 tune. The comparison to real data collected by the CMS experiment indicates that the agreement is much better for the excited states than for the ground state. The observed fast increase of the \meanptups{} at small values of \ntrack{} is mainly due to the contribution from the away region. Furthermore, when computing the \meanptups{} from jetty and isotropic events a clear \pt{} hardening is observed in jetty events. Finally, analyzing the fragmentation of jets containing an \ups{} it is proposed a new method to test the new quarkonia shower present in the Monte Carlo event generator.


\end{abstract}


\end{frontmatter}


\section{Introduction} 

The November Revolution is the period in time characterized by the discovery of the J/$\psi$ particle \citep{JpsiDiscovery1,JpsiDiscovery2}. As an aftermath, new heavy hadrons containing the charm quark were experimentally detected. Three years later a new heavy meson, called $\Upsilon$, was discovered and it was rapidly identified as a bound state of a new heavy quark: the bottom or beauty \citep{Ups1SDiscovery}. The top quark measured by the first time in 1995, has such a short lifetime that it can not form stable bound states \citep{Top1,Top2}. However recent results from LHC experiments have demonstrated an excess of a top-antitop production in an invariant mass region consistent with the pair production threshold \citep{TopAntiTop1,TopAntiTop2}.

Bound states of a heavy quark and its coresponding antiquark are commonly known as quarkonia. Quantum Chromodynamics (QCD), following the assumption of factorization, dictates that quarkonia production in proton-proton collisions is an admixture of both perturbative and non-pertubative QCD \citep{FactorizationTheo}. The creation of the heavy quark-antiquark pair (\qqbar) due to the hard partonic interaction is within the regime of pertubative QCD, while the hadronization process into a defined quarkonium state is part of the non-perturbative QCD domain. As a consequence, this last step relays on phenomenological models.

On one hand the Color Singlet Model assumes that the \qqbar{} pair created in the partonic scattering is produced as a color singlet state \citep{CSM0,CSM1,CSM2}. On the other hand the Color Evaporation Model makes no such assumption, so the \qqbar{} pair can be in a color singlet or color octet state \citep{CEM1,CEM2,CEM3}. In the latter case the color is neutralized by gluon emission.

The Non-Relativistic QCD (NRQCD) model assumes that, due to the large masses of the quarks they can be treated with a non-relativistic formalism \citep{NRQCD1,NRQCD2}. In this model the process that describes the evolution from a \qqbar{} into a physical quarkonium is governed by Long Distance Matrix Elements (LDME).

Quarkonia has extensively been studied at the LHC. However, in recent years quarkonia production in proton-proton (pp) collisions as a function of the multiplicity has drawn important attention. The reason is that very little is known about the interplay between hard (\qqbar{} production) and soft (underlying event) processes. As a consequence, global event observables may vary depending on the quarkonium state produced. For \ups, with quantum number $n$ = 1, 2 and 3, the ALICE experiment has shown that the increase of the self-normalized yields is compatible with a linear scaling of the charged particle multiplicity, confirming the Multiple Partonic Interaction (MPI) scenario implemented in Pythia 8  \citep{ALICEups}. The ATLAS experiment’s observations indicate an important difference in the charged particle multiplicity in events where different \ups{} are produced, something that is not properly reproduced by Pythia 8, implying that the connection between hard and soft processes is still not well modeled \citep{ATLASups}. CMS and LHCb have measured the production cross section ratios of the excited to ground $\Upsilon$ state, as a function of the multiplicity \citep{CMSupsMult,LHCbUps}. Both experiments report a decreasing trend, indicating that final state effects can strongly affect the \ups{} states.

There are also recent theoretical advances intended to explain the J/$\psi$ production as a function of the multiplicity in pp collisions. A formulation based on BFKL Pomeron calculation assumes that the quarkonia state is produced by a triple gluon fusion and that the multiplicity distribution is close to a negative binomial when a large number of particles are produced \citep{HighEnergyQCD}. Other models rely on the Color Glass Condensate framework to explain the faster than linear increase in the experimental measurements as an effect of the saturation \citep{EventEngineering,Fluctuations}. 

In the context of heavy-ion physics it was theorized that the presence of the Quark Gluon Plasma (QGP) generates a sequential suppression on the \ups{}, something that was recently verified by CMS \citep{FeedDown,UpsilonSup}. Furthermore, since the start of the LHC heavy-ion like phenomena have been measured in high multiplicity pp and proton-lead collisions \citep{Collectivitypp,CollectivitypPb}. This has triggered the debate if these effects indicate the presence of a mini-QGP or it is the result of some other mechanism \citep{MiniQGP,FlowLike}. In this sense, studying the \ups{} production and kinematics in small collision systems with a large number of charged particles produced could shed some light on such controversy.

\section{Color reconnection}

Pythia 8 is a Monte Carlo event generator widely used to describe high energy pp collisions at the LHC \citep{Pythia8.3}. In general, the event generator divides a pp collision in two parts. The first one is related to the hard partonic scattering, characterized by a large transverse momentum exchange. The second is the underlying event (UE) and comprises the additional activity that includes MPI, beam remnants and initial and final state radiation. 

In the partonic scatterings, quarks and gluons are connected by color strings, creating a dense net of color lines. The original leading color approximation consisted of an infinite number of colors, such a way that generated partons could only be connected to their emitting parents. The actual Color Reconnection (CR) model allows color lines among partons created in different MPI, implying a more complex color topology where the possible color reconnections are those that minimize the string length. Indeed, the reconnection probability is set to be higher for soft MPI systems as a result of their extended wave function that can easily overlap with other MPI.  As it is not possible to have a description of an hadronic collision fully extracted from theoretical foundations, the different CR models contain free parameters. The MPI-based CR is the simplest one as it only contains the Reconnection Range (RR) as tuneable parameter. The model sets the probability of a low \pt{} parton to be merged with a higher \pt{} one via the RR. It is well known that variations on the RR value have an important effect on the inclusive charged particle production and transverse momentum. Indeed, RR values close to zero lead to non-physical results, while very little modification of the distributions are obtained for RR $>$ 3 \citep{CReffects}.

\begin{figure*}
\begin{center}
\includegraphics[width=0.98\textwidth]{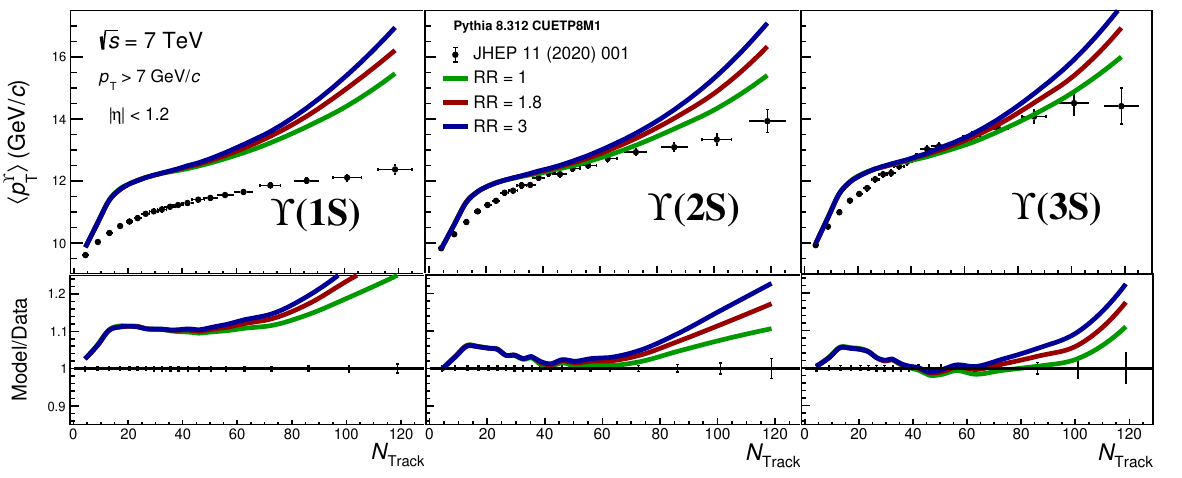}
\caption{Black dots are the \ups{} mean transverse momentum (\meanptups) as a function of the charged particle multiplicity (\ntrack) for pp collisions at \s{} = 7 TeV measured by the CMS experiment \citep{CMSupsMult}. The vertical error bars represent the quadratic sum of the statistical and systematic uncertainties, while horizontal bars are the uncertainties on \ntrack. Full lines are the predictions from Pythia 8.312 CUETP8M1 (red) and other configurations where the only variation is the value of the color reconnection range: $RR$ = 1 (green) and $RR$ = 3 (blue). Bottom panel presents the model to data ratio where the vertical lines at the unity is the data error.}
\label{PublishedPlots}
\end{center}
\end{figure*}

Unless indicated otherwise, the following study employs Pythia 8.312 together with the CUETP8M1 tune, the MPI-based CR and the default value of RR = 1.8 \citep{Pythia8.3,TuneCUETP}. From now onwards this configuration will be simply called Pythia 8.312 CUETP8M1. In figure \ref{PublishedPlots} black dots are the \ups{} mean transverse momentum (\meanptups) as a function of the charged particle multiplicity (\ntrack) reported by the CMS experiment for pp collisions at \s{} = 7 TeV \citep{CMSupsMult}. The vertical error bars represent the quadratic sum of the statistical and systematic uncertainties, while horizontal bars are the uncertainties on \ntrack{}. The measurement is performed for \ups{} with \pt{} $>$ 7 \gevc{}, $|y| <$  1.2 and decaying to muons with $|\eta| <$ 2.4. Furthermore, each muon must survive an $\eta$-dependent \pt{} requirement. The charged particle multiplicity is defined by the number of reconstructed tracks with \pt{} $>$ 0.4 \gevc{} and $|\eta| <$ 2.4. Notice muon tracks are not taken into account for \ntrack{}. The experimental measurement is compared to \py{} CUETP8M1 (red) and to other configurations where the only variation is the value of the color reconnection range: $RR$ = 1 (green) and $RR$ = 3 (blue). Qualitatively, the \meanptups{} of the three \py{} configurations for the three resonances show the same behavior: a steep increase in the low multiplicity regimen followed by a slower growth at mid multiplicity and finally another fast rise for large \ntrack{}. So, CR has little effect on the description of the data. The agreement between simulations and data is better for the excited states and the variation of the color reconnection range only becomes important at very high multiplicities. According to the bottom panel of figure \ref{PublishedPlots}, for $\Upsilon$(1S) the model to data ratio stays well below the 15$\%$ difference for \ntrack{} $\lesssim$ 80. Above this value, the simulations completely overshoot the data. However, less than 3$\%$ of the total events are found to produce \ntrack{} $>$ 80. For $\Upsilon$(2S) and (3S) there is less than 10$\%$ relative difference between data and simulation for \ntrack{} $<$ 90 and 100, respectively. 

On one hand the simulation shows that the \meanptups{} distribution is basically the same for the three resonances, except by a minimal hardening given by the mass increase on the radially excited states. On the other hand, the data show a notable increase of the \meanptups{} from the ground to the excited states. It is due to this difference that there is a better agreement between data and simulation for $\Upsilon$(2S) and (3S). However, an improvement of NRQCD is required to better describe the \meanptups{} as a function of the multiplicity.

\begin{figure*}
\begin{center}
\includegraphics[width=0.65\textwidth]{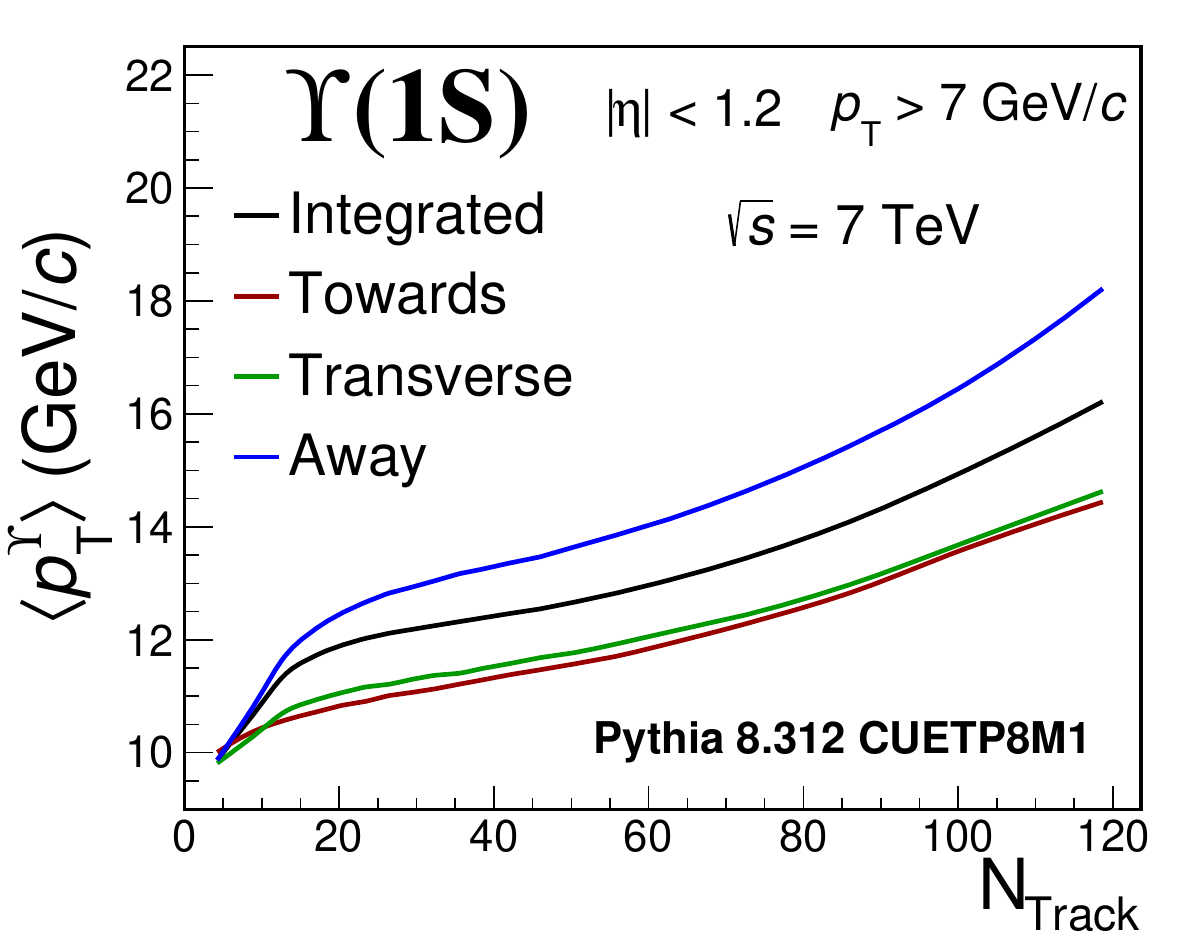}
\caption{$\Upsilon$(1S) \meanptups as a function of \ntrack for three different azimuthal regions relative to the leading charged particle: towards (red), transverse (green) and away (blue). The integrated (black) distribution is included for comparison.}
\label{ThreeRegions}
\end{center}
\end{figure*}

Additional insight can be obtained by computing the \meanptups{} in three different azimuthal regions relative to the leading charged particle (the one with the largest \pt{} of the event). The azimuthal angle ($\phi$) is located in the plane perpendicular to the beam axis and the azimuthal angular difference ($\Delta\phi$) between the leading particle and the \ups{} is used to define three regions: towards ($|\Delta\phi|$ $<$ $\pi/3$), transverse ( $\pi /3$ $<$ $|\Delta\phi|$ $<$ $2\pi /3$) and away ($|\Delta\phi|$ $>$ $2\pi /3$). Figure \ref{ThreeRegions} shows the $\Upsilon$(1S) \meanptups{} in the towards (red), transverse (green) and away (blue) regions compared to the integrated one (black). As can be seen, the steep increase of the \meanptups{} at low multiplicities is basically due to the away region. At large values of \ntrack{} the three components present a fast rise, albeit the largest contribution is once again from the away region. Similar results are obtained for the excited states.

\section{Spherocity}

Event shape observables are commonly used to separate the different geometrical and topological configurations of the final state particles produced in high energy particle collisions. As these observables (thrust, broadening, sphericity, etc) are infrared and collinear safe they have been extensively used to tune the parton shower and non-perturbative elements in a variety of Monte Carlo event generators \citep{EventShapes}. 

Among the event shape variables, transverse spherocity (from now onwards called spherocity), is defined relative to a unit vector ($\hat{n}$) that minimizes the ratio \citep{Spherocity}:

\begin{equation}
S_0 = \frac{\pi^2}{4} \left(  \frac{ \sum_i \mid \vec{p}_{\mathrm{Ti}} \times \hat{n} \mid}{\sum_i p_{\mathrm{Ti}}} \right)^2.
\end{equation}

The sum index runs over all charged particles that survive the track selection mentioned in previous section. This is a normalized observable, implying the range is between 0 and 1. Spherocity is a very useful quantity as it provides valuable information on the final state azimuthal topology of the collision. Indeed, events in the lower edge of $S_0$ (0-0.1) are characterized by a back-to-back, pencil-like, structure and are called jetty events. On the opposite side, the isotropic events are those with $S_0$ close to unity (0.9-1) and indicate a symmetric distribution of particles. Isotropic events are the outcome of soft processes, while the jetty ones are the result of hard events. In Pythia 8 isotropic events are associated with a high underlying event activity and, as a consequence, with a large number of MPI. The UE activity decreases when spherocity is reduced. 

\begin{figure*}
\begin{center}
\includegraphics[width=0.49\textwidth]{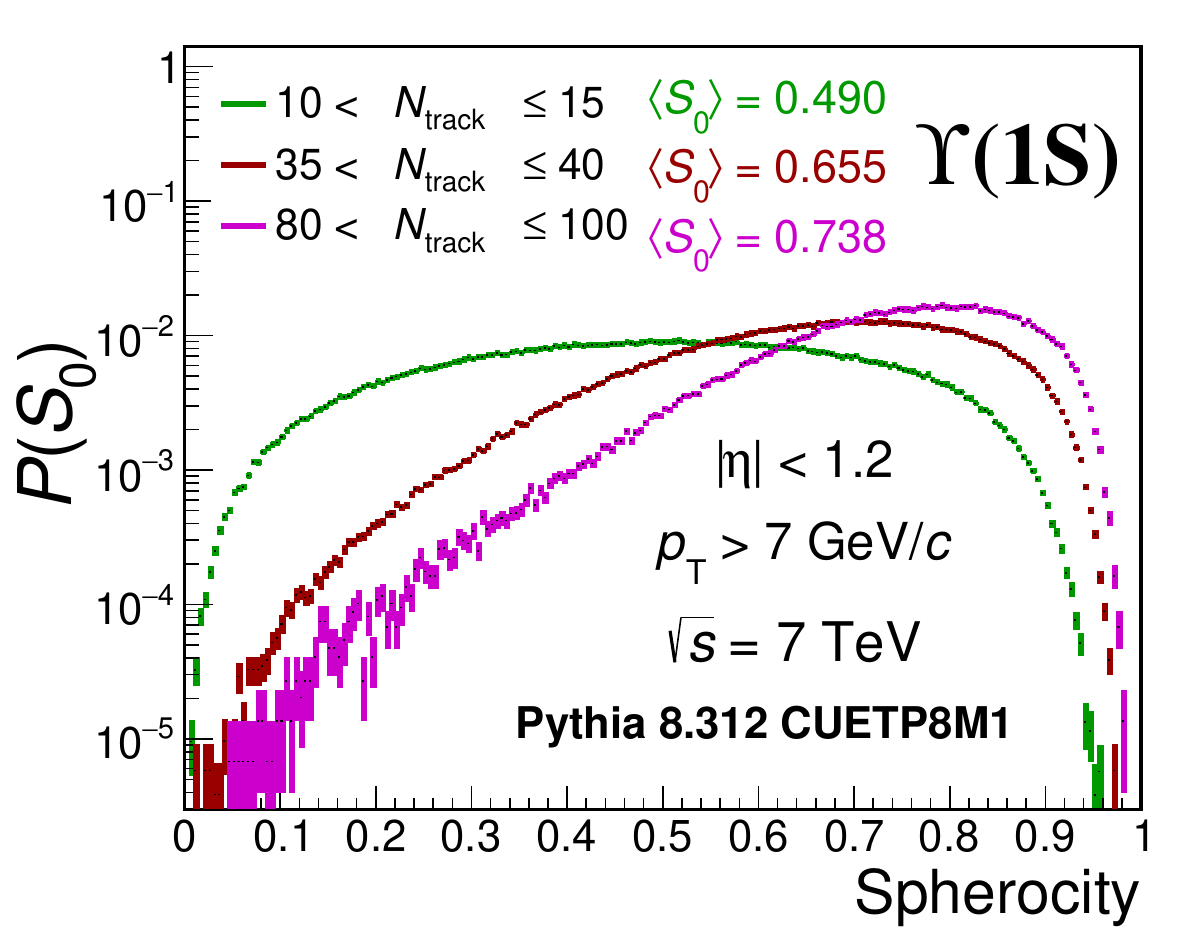}
\hspace{-0.4cm}
\includegraphics[width=0.49\textwidth]{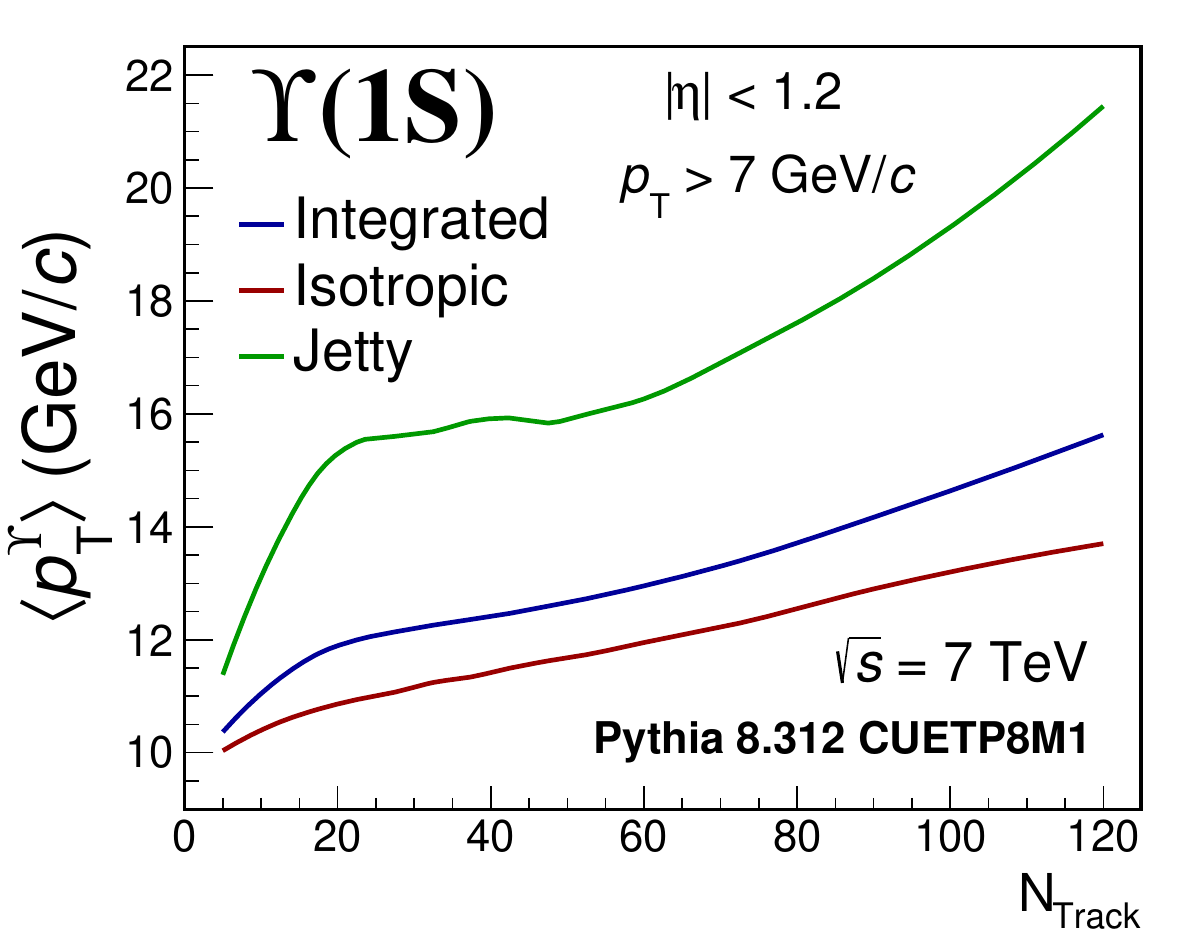}
\caption{Left: spherocity distributions for $\Upsilon$(1S) only. The plot presents the distributions for low (green), mid (red) and high (magenta) multiplicities with their corresponding mean values. Right: $\Upsilon$(1S) \meanptups for isotropic (red) and jetty (green) events together with the $S_0$ integrated (blue).}
\label{FigSpherocity}
\end{center}
\end{figure*}

Figure \ref{FigSpherocity} (left) shows some spherocity distributions for $\Upsilon$(1S) only. Specifically, the plot presents the corresponding distributions for low (green), mid (red) and high (magenta) multiplicities. For quantitative comparisons, also the mean values of $S_0$ are displayed. The $S_0$ curves are computed with the kinematics of the tracks used to assess the multiplicity of the events. The figure clearly indicates that the $S_0$ curves are depleted towards larger spherocity values as \ntrack{} increases. This means that for large multiplicity events the probability to have an event with spherocity close to unity grows, so these events are more likely to be classified as isotropic. Same behavior is found for $S_0$ distributions corresponding to $\Upsilon$(2S) and $\Upsilon$(3S).

Figure \ref{FigSpherocity} (right) displays the $\Upsilon$(1S) \meanptups{} for isotropic (red) and jetty (green) events together with the $S_0$ integrated (blue). On one hand the mean transverse momentum extracted from jetty events presents a very fast increase for \ntrack{} $\lesssim$ 20, then an almost flat behavior for 20 $\lesssim$ \ntrack{} $\lesssim$ 50 and finally a rapid rise with the multiplicity. On the other hand, the \meanptups{} for $\Upsilon$(1S) in isotropic events depicts a slow growth for all the \ntrack{} range. The relative difference between the jetty and the spherocity integrated distributions reaches up to 40$\%$ for events with the largest multiplicities. Such a large difference implies a clear \pt{} hardening for jetty events. Furthermore, this plot indicates that the $S_0$ integrated distribution is mainly constituted by a group of isotropic topologies where jet-like events are not that common. These observations have already been measured by the ALICE experiment for lighter particles using an unweighted transverse spherocity, a variant of the $S_0$ used in this document \citep{S0forLF}. Similar results are found for $\Upsilon$(2S) and $\Upsilon$(3S).

\section{\ups in jets}

Both CMS and LHCb experiments have studied the fragmentation of jets containing prompt or non-prompt J/$\psi$ in pp collisions at \s{} = 5.02 and 13 Tev, respectively \citep{JpsiInJetsCMS,JpsiInJetsLHCb}. The results are expressed as a function of $z$, the jet transverse momentum fraction carried by the J/$\psi$. The comparison to Pythia 8 for prompt J/$\psi$ shows that it overestimates the data when $z \approx 1$, indicating that the event generator predicts that prompt J/$\psi$ are mainly produced with a small degree of surrounding jet activity, a different trend observed from the data.

In a recent publication it is shown that a better agreement between data and the event generator can be achieved by activating the new quarkonia parton shower available in recent Pythia 8 versions \citep{JpsiUpsilonJets}. This new implementation further enhances the leading order NRQCD based prediction available in Pythia 8 for quarkonia production by introducing quarkonia splittings during the parton shower \citep{OniaShower}. As part of the results, the publication shows a prediction for $\Upsilon$(1S) production in jets with 30 $<$ \pt{} $<$ 40 \gevc{}. Opposite to the prompt J/$\psi$ case, for $\Upsilon$(1S) the activation of the new quarkonia shower yields the same results as when it is not initialized. The explanation for this effect relies on the mass difference between the J/$\psi$ and the $\Upsilon$(1S), being the latter more massive requires harder partons to create an $\Upsilon$(1S) during the shower. The conclusion is that the outcome of the new quarkonia shower for $\Upsilon$(1S) is only visible for high \pt{} jets. 

\begin{figure*}
\begin{center}
\includegraphics[width=0.98\textwidth]{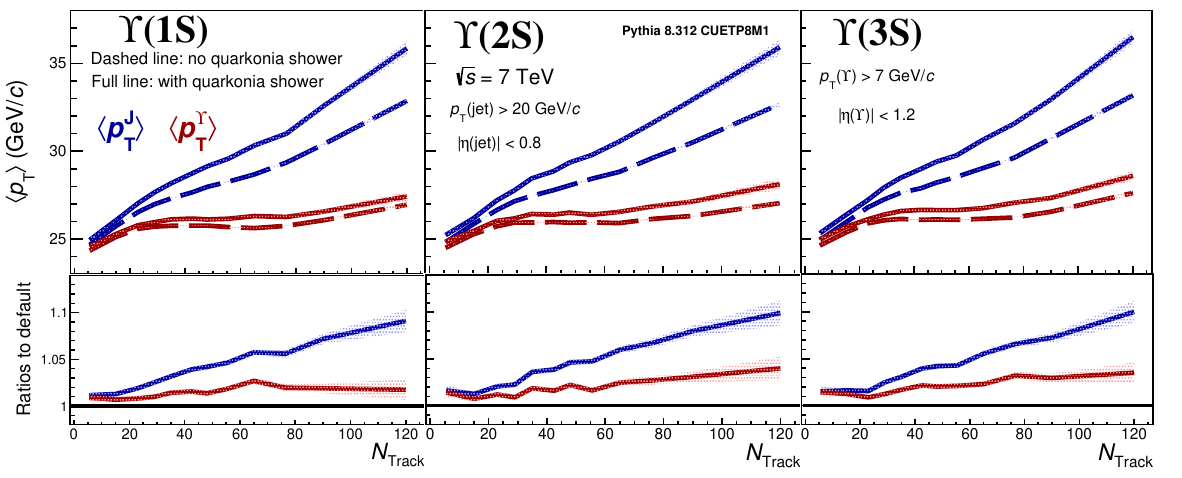}
\caption{Mean transverse momentum of jets $\langle p_{\mathrm{T}}^{\mathrm{J}} \rangle$ (blue) and \ups $\langle p_{\mathrm{T}}^{\Upsilon} \rangle$ (red) computed with (full line) and without (dashed line) the new quarkonia shower activated. The bottom panel shows the ratios of \meanptups and \meanptjet when the new quarkonia shower is activated relative to the case when it is not (default configuration). Shaded areas indicate the statistical uncertainties.}
\label{FigJets}
\end{center}
\end{figure*}

In this section the mean transverse momentum of jets (\meanptjet) and \meanptups{} as a function of \ntrack{} is used to provide an alternative approach to test the new quarkonia shower with low \pt{} jets. Figure \ref{FigJets} shows the $\langle p_{\mathrm{T}}^{\mathrm{J}} \rangle$ (blue) and $\langle p_{\mathrm{T}}^{\Upsilon} \rangle$ (red) computed with (full line) and without (dashed line) the new quarkonia shower activated. The anti-$k_{\mathrm{T}}$ algorithm from FastJet is employed for the jet clustering, demanding a radius $R =$ 0.4, \pt{} (jet) $>$ 20 \gevc{} and limiting the axis to $|\eta \mathrm{(jet)}| <$ 0.8 \citep{FastJet}. In this way the pseudorapidity requirement of the \ups{} ($|\eta (\Upsilon)| <$ 1.2) is consistent with the full jet cone (axis and radius). All selected jets contain an \ups{} as it, not the daughter muons, is used as direct input to the jet clustering algorithm and as a consequence \meanptups{} $\lesssim$ \meanptjet{}. Indeed, at low multiplicities it becomes highly probable for jets to be solely composed by an \ups{} so \meanptups{} $\simeq$ \meanptjet{}. For large values of \ntrack{} such possibility vanishes as there are more particles to be clustered in the jet so \meanptups{} $<$ \meanptjet{}. The \meanptjet{} distributions are monotonically increasing from low to high multiplicity. A somehow different behavior is found for the \meanptups{}, where there is no big variation in the whole \ntrack{} range. As can be seen, including the new quarkonia shower hardens the mean transverse momentum distribution of the \ups{} and the jet itself. This is a natural consequence of the inclusion of the new quarkonia splittings during the parton shower. Such effect is somehow hidden when simply using the $z$ variable as in \citep{JpsiUpsilonJets}. The bottom panel of figure \ref{FigJets} shows the ratios of \meanptups{} and \meanptjet{} when the new quarkonia shower is activated relative to the case when it is not (default configuration), indicating that the effect is more notorious as \ntrack{} increases. So, the new quarkonia shower implementation can be tested by computing \meanptups{} and \meanptjet{} from the event generator and compare it to real data.

\section{Conclusion}

A study on the multiplicity dependence of the $\Upsilon$(nS) mean transverse momentum in proton-proton collisions at \s{} = 7 TeV has been performed using \py{} CUETP8M1 tune. The description of the experimental results by the event generator is much better for the excited states than for the ground state. By computing the \meanptups{} in the towards, transverse and away azimuthal regions (relative to the leading charged particle) it is observed that the steep increase of the distribution at low multiplicities comes from the away region. Furthermore, there is a clear hardening of the \meanptups{} from jetty events relative to the isotropic ones, indicating that the integrated spectra is mainly constituted by a group of isotropic topologies where jet-like events are not that common. Finally it has been shown that it will be possible to distinguish among predictions including or not the new quarkonia shower using low \pt{} jets containing an \ups{}. This new propossal shows that the best is to compare the \meanptjet{} distributions as a function of \ntrack{} from the event generator to real data, as in this case the relative difference between the two shower configurations increases with the multiplicity reaching values slightly larger than 10$\%$.

\section{Acknowledgments}

This work has employed an important amount of computing resources from the ACARUS at the Universidad de Sonora, without this facility it would have been impossible to develop this study.


\bibliography{mybibfile}

\end{document}